# Effect of particle-momentum on an isothermal flow-field inside a swirl combustor

Madan Lal Mahato and Nitesh Kumar Sahu*

Department of Fuel, Minerals and Metallurgical Engineering, Indian Institute of Technology (ISM) Dhanbad, Jharkhand-826004, India.
*Corresponding author: nitesh@iitism.ac.in

**ABSTRACT**

This paper investigates the impact of particles on isothermal flow inside a lab-scale swirl combustor for a fixed inlet swirl number of 0.67 using steady-state CFD simulations. The combustor geometry and baseline conditions, with no particles, are taken from Taamallah *et al.* [1], but with a simplification. In the present work, we provide rotation to the flow using velocity boundary condition, whereas in [1], a swirler is built into the geometry to achieve the same effect. Shear stress transport (SST) $k$–$\omega$ model, an eddy-viscosity based Reynolds averaged Navier-stokes equation approach, is used for modelling turbulence. The comprehensive model is validated against the experimental axial-velocity data in [1]. Two simulations, one with 75 and another with 100 μm particles using Discrete particle model (DPM) were conducted to isolate the effect of particle motion on swirl-combustor flow without combustion. Their analysis shows significant downstream shift of central recirculation zone (CRZ). An effect that can significantly impact the stabilization of coal flame in pulverized particle reactors.

**Keywords**: Swirl combustor; central recirculation zone; discrete particle model; isothermal flow.

## 1. INTRODUCTION

Swirl-stabilized flows are widely used in gas turbine, industrial burners and furnaces to enhance flame anchoring and mixing. A prominent feature of such flows is the formation of a central recirculation zone (CRZ), a low-velocity region on the axis that recirculates hot gases and active species back upstream [2]. In practice the CRZ helps stabilize flames and extends lean blow-off limits [3]. In entrained flow reactors, CRZ increases residence time of particles thereby enhancing its performance [4]. However, most studies focus on this flow structure in a reacting environment, which is known to modulate the flow field [1,3,5,6]. Therefore, understanding the CRZ formation even in non-reacting flows and dispersed-phase isothermal flows is important for combustor design.

## 2. LITERATURE REVIEW AND OBJECTIVE

Several researchers have contributed to understanding the behaviour of the central recirculation zone (CRZ) in swirling flows. Syred [7] identified a critical swirl number of approximately 0.6, beyond which vortex breakdown occurs in reacting gaseous flows. Chakroun [8] in a gas-phase reacting environment examined the influence of equivalence ratio and found that increasing it leads to a reduction in the size of CRZ. Sahu *et al.* [4] observed the impact of mass flow rate and the pressure on the CRZ in an entrained flow gasifier. They reported decrease in the size of CRZ upon increasing the two investigated parameter. Sahu *et al.* [9] investigated the impact of pulverized particle size on CRZ in an entrained flow gasifier, where the CRZ size decreased with the particle size found that changing the particle size. However, the flow field is known to be modulated by a reacting environment therefore to isolate the effect of particle size solely we perform this study since relevant investigations are scarce.

## 3. METHODOLOGY

### 3.1 Computational Modelling

We develop our computational model referring to the lab-scale swirl combustor geometry of Taamallah *et al.* [1], shown in Figure 1. It features a straight inlet pipe followed by a sudden expansion into a cylindrical chamber, where rotation is imparted to the flow at the inlet using a velocity boundary condition. The inlet section has a diameter of 0.038 m and length of 0.045 m, while the main chamber has a diameter of 0.076 m and length of 0.225 m. The base flow conditions are specified in Table 1. The inlet Reynolds number of the flow is ~20,000. The flow field is modeled as a steady, incompressible, isothermal flow using Reynolds-averaged Navier–Stokes equations approach, with shear-stress-transport (SST) $k$–$\omega$ model for turbulence closure. The simulation is performed using ANSYS Fluent 2024R2 commercial software.

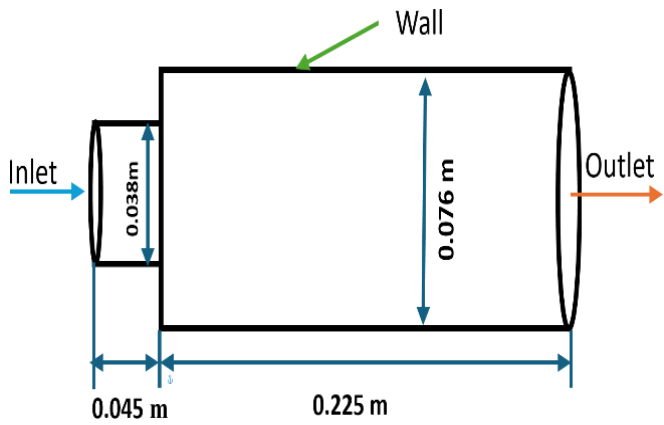

Figure 1: Schematic of the combustor geometry (Taamallah *et al.* [1]).

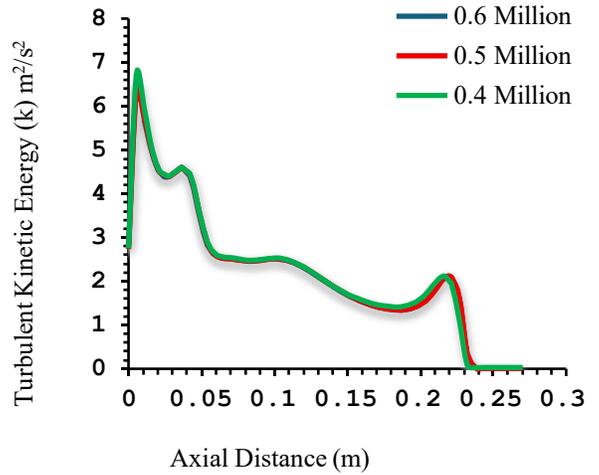

Figure 2: Grid independence study

Table 1: Operating conditions of the base case [1]

| Velocity (m/s) | Pressure (bar) | Temperature (K) |
|---|---|---|
| 8.2 | 1.0 | 300 |

### 3.2 Boundary Conditions

A uniform velocity profile with a magnitude of 8.2 m/s is imposed at the inlet using velocity inlet boundary condition (b.c.). The outlet is set to ambient pressure using pressure outlet boundary condition. No-slip b.c. is imposed at walls.

### 3.3 Solver Settings

We use pressure-based steady-state coupled solver with second-order upwind discretization scheme for convective terms and central difference scheme for the diffusion terms. PRESTO scheme is used for pressure interpolation. The scaled residuals of all the quantities quickly reduced below $\sim 10^{-6}$. The convergence was confirmed by monitoring the axial- velocity over iterations at the two points within the domain.

### 3.4 Grid Independence Study

A grid independence study was performed by comparing the axial-variation of turbulent kinetic energy ($k$) on three successively refined meshes, shown in Figure 2. All elements in these meshes are hexahedral. The change in plotted variation between the two finer meshes is insignificant, so we select the grid with 0.5 million hexahedral elements for our investigations. The orthogonal quality and aspect ratio of the selected grid, shown Figure 3, are 0.52 and 6.7, respectively. The $y^+$ of the first grid elements from the wall is around 20.

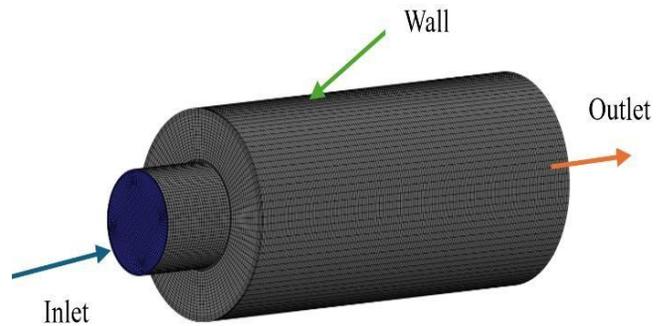

Figure 3: Computational mesh with 0.5 million hexahedral elements.

### 3.5 Validation Study of Base Flow and Discrete Particle Model

We compare the computed radial variations of axial- velocity profiles with the experimental data in [1]. The comparison at two axial locations x/R = 0 and 0.5, where x corresponds to distance from the expansion and R= 0.038 m, is shown in Figure 4. The agreement between the computed and experimental results seems reasonable. Hence, we extend the model for additional investigations.

Discrete particle model (DPM) is added to the base flow model to investigate the impact of pulverized particles, where the volume fraction of particles is less than 1%, a typical behavior of coal reactors [10]. DPM is a well-known to simulate the effect of dispersed phase on fluid flow [11]. The interaction between particles and gas flow is accounted via two- way coupling. Two new cases, one with discrete particles of diameter 70 $\mu$m and another with discrete particles of diameter 100 $\mu$m are simulated using steady DPM model approach. The particles having a density of 1000 kg.m³ are injected at the rate of 0.093 kg/s, based on less than 1% volume fraction. 6000 particles are tracked in each simulation, where the effect of turbulent dispersion is considered using Discrete random walk model. The impact of particles on the flow field in discussed below.

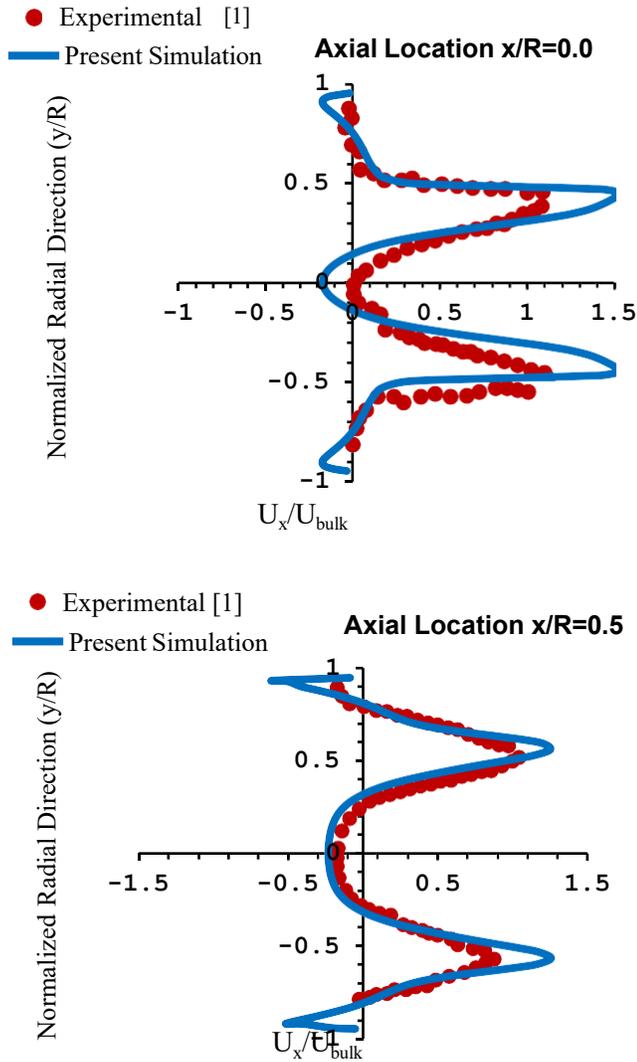

Figure 4: Comparison of radial profiles of axial-velocity with experimental data [1] at x/R = 0.0 and 0.5.

## 4. RESULTS AND DISCUSSION

The influence of particle on CRZ can be seen in the contours of axial-velocity for the three simulated cases in figure 5. A significant downstream shift in the location of CRZ is visible in both the particle-laden cases compared to the baseline case with no particles. This shift can be attributed to the momentum imparted by particles to the flow.

Figure 6 shows the axial-variation of axial velocity within the combustor. The presence of particles strengthens the intensity of the negative axial velocity in the CRZ but shifts it downstream. The 100 μm particles induce a larger shift than the 75 μm ones, indicating that larger particles with higher inertia more significantly influence the CRZ position. However, the vortex breakdown details are not studied here, as the focus remains on CRZ displacement. A shift of 0.17 m is observed in 100 μm particle case and 0.145 m in the 75 μm particle case compared to the base case.

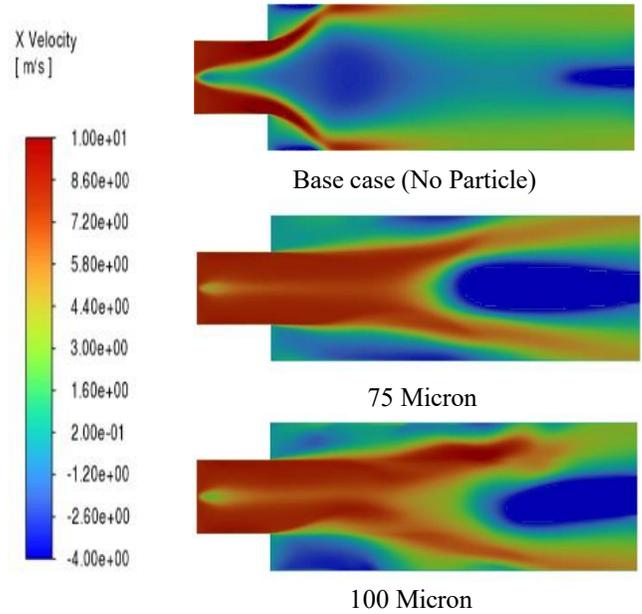

Figure 5: Contour of axial-velocity.

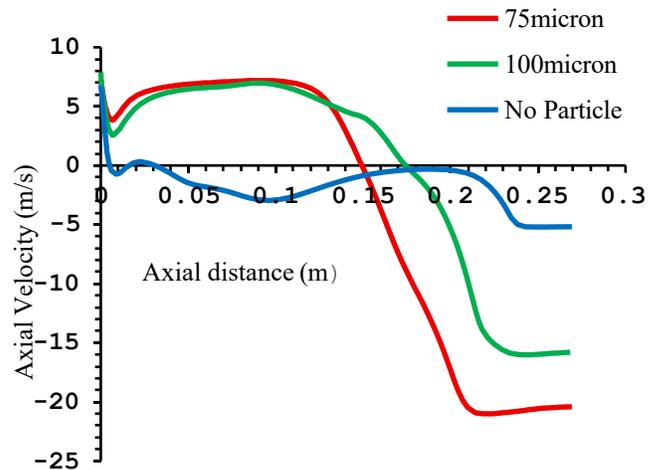

Figure 6: Axial variation of axial velocity.

## 5. CONCLUSIONS

The present work numerically investigates the impact of particle on the central recirculation zone formed in an isothermal swirling flow inside a lab-scale combustor. The flow has an inlet swirl number of 0.67. A comprehensive model with SST k-w turbulence closure model and discrete particle sub- model for particle interaction and other sub-models is developed. The analysis of the simulated results reveals significant shift in the position of CRZ compared to the case with no particles. The shift is also impacted by the size of particles. These findings are particularly relevant for pulverized fuel combustor designs, where particle momentum can alter flame stabilization location

## ACKNOWLEDGEMENTS

The present work is supported by the Faculty Research Scheme (FRS) Grant; FRS (217)/2024-25/FMME of IIT-ISM Dhanbad.

## NOMENCLATURE

| | | |
|---|---|---|
| *CRZ* | Central Recirculation Zone | -- |
| *DPM* | Discrete Particle Model | -- |
| *SST k–ω* | Shear Stress Transport *k–ω* Model | -- |
| *Re* | Reynolds Number | -- |
| *ρ* | Density | [kg/m$^3$] |
| *S* | Swirl Number | -- |
| *U* | Axial Velocity | m/s |
| *CFD* | Computaional Fluid Dynamics | -- |
| *k* | Turbulent Kinetic Energy | m$^2$/s$^2$ |